\documentclass[sigconf,nonacm]{acmart}
\AtBeginDocument{%
  \providecommand\BibTeX{{%
    \normalfont B\kern-0.5em{\scshape i\kern-0.25em b}\kern-0.8em\TeX}}}

\acmConference[JCDL '23]{2023 ACM/IEEE Joint Conference on Digital Libraries (JCDL)}{June 26--30,
  2023}{Santa Fe, New Mexico, USA}
  
\usepackage{booktabs} % For formal tables
\usepackage[toc,page]{appendix}
\usepackage{subcaption}
\usepackage[inline]{enumitem}
\usepackage{graphicx}
\usepackage{verbatim}
\usepackage{svg}
\usepackage{xcolor}
\usepackage{float}
\usepackage{dblfloatfix}

\begin{document}

\title[From Speech to Search]{Toward Connecting Speech Acts and Search Actions in Conversational Search Tasks}

\author{Souvick Ghosh}
\email{souvick.ghosh@sjsu.edu}
\orcid{0000-0003-1610-9038}
\affiliation{%School of Information, 
  \institution{San José State University}
  \streetaddress{One Washington Square}
  \city{San José}
  \state{CA}
  \country{USA}
  \postcode{95192-0029}
}

\author{Satanu Ghosh}
\email{satanu.ghosh@unh.edu}
\orcid{0000-0003-4232-9931}
\affiliation{%Department of Computer Science,
  \institution{University of New Hampshire}
  \streetaddress{33 Academic Way, Kingsbury Hall, N229}
  \city{Durham}
  \state{NH}
  \country{USA}
  \postcode{03824-2619}
}

\author{Chirag Shah}
\email{chirags@uw.edu}
\orcid{0000-0002-3797-4293}
\affiliation{
  \institution{University of Washington}
  \streetaddress{Mary Gates Hall 330Q}
  \city{Seattle} 
  \state{WA} 
  \country{USA}
  \postcode{98195-2840}
}

%\author{Anonymous Authors}

%^\renewcommand{\shortauthors}{Gh and Tobin, et al.}

\begin{abstract}
Conversational search systems can improve user experience in digital libraries by facilitating a natural and intuitive way to interact with library content. However, most conversational search systems are limited to performing simple tasks and controlling smart devices. Therefore, there is a need for systems that can accurately understand the user's information requirements and perform the appropriate search activity. Prior research on intelligent systems suggested that it is possible to comprehend the functional aspect of discourse (search intent) by identifying the speech acts in user dialogues. In this work, we automatically identify the speech acts associated with spoken utterances and use them to predict the system-level search actions. First, we conducted a Wizard-of-Oz study to collect data from 75 search sessions. We performed thematic analysis to curate a gold standard dataset -- containing 1,834 utterances and 509 system actions -- of human-system interactions in three information-seeking scenarios. Next, we developed attention-based deep neural networks to understand natural language and predict speech acts. Then, the speech acts were fed to the model to predict the corresponding system-level search actions. We also annotated a second dataset to validate our results. For the two datasets, the best-performing classification model achieved maximum accuracy of 90.2\% and 72.7\% for speech act classification and 58.8\% and 61.1\%, respectively, for search act classification. \vspace*{-1em}
\end{abstract}




\maketitle

\section{Introduction}
Digital libraries comprise vast digital collections of data, which can be challenging for users to navigate and access~\citep{machidon2020culturalerica}. An interactive intelligent system can greatly assist users in navigating the data, fulfilling their information needs, and enhancing their search and interaction experience~\citep{machidon2020culturalerica, mckie2019enhancing}. 
In traditional library systems, users -- unsure of their needs~\citep{tuominen1997user} -- engage in a back-and-forth discourse with the librarian, who helps the user to formalize their information need into queries. The search are often complex and non-factoid in nature and multiple documents may be accessed to obtain the correct and most helpful information. 
Similarly, in digital libraries, users' questions may be non-factoid, necessitating a proactive intermediation strategy that can be supported by an intelligent conversational system~\citep{crabtree1997talking, brewer1996role}. 
Therefore, in this study, we investigate how users interact with an intelligent system when their information needs are non-factual in nature. We draw connections between the speech or dialogue acts (from user utterances) and system-level actions.
The framework presented in this paper can be applied to an intelligent conversational agent for digital libraries.

Over the past decade, information retrieval systems have made significant advancements. These systems, once limited to textual retrieval, have evolved to facilitate multi-modal retrieval, enabling users to interact more naturally and intuitively with the system. Consequently, the primary focus of search systems has transitioned from merely organizing and ranking documents to making them more user-friendly and interactive.
With the proliferation of smarter mobile devices, it has become possible to use conversational search systems (chatbots and personal assistants) for everyday tasks.

However, mobile computing devices like smartphones or smart wearables come with limited display capabilities; therefore, typing or reading is not a pleasant experience for the users~\citep{yankelovich1995designing}. In contrast, voice-based interactions are more user-friendly and convenient when using 
these devices~\citep{chang2002system,najjar1998user,%trippas2015results,
turunen2012evaluation}. Personal assistants in our smartphones have proven to be useful companions as they allow us to perform simple tasks like setting the alarm or making a call quickly while doing other daily chores~\citep{DBLP:conf/aiia/FrummetEL19,frummet2022can,ghosh2019informing,guy2016searching}.  
Conversational search systems can also aid people with visual impairment and learning disabilities in reading~\citep{sahib2012accessible, sahib2012comparative, LaiJ2006SID}. However, the use cases of personal assistants are quite limited currently, and they are incapable of comprehending complex queries~\citep{luger2016like}. Also, they cannot engage in longer conversations or maintain context over multiple turns~\cite{luger2016like}, which makes widespread adoption of such systems difficult.

In human-human conversations, multiple turns help the participants build context and clarify doubts~\citep{radlinski2017theoretical, aliannejadi2019asking, zamani2020generating}. Since user utterances represent the search intent, they should be used as guidelines for query formulation, search action, and relevance assessment. Also, a combined understanding of speech acts and search actions can help us recognize topic changes during conversational search sessions~\citep{hienert2019recognizing}, which is an important direction in this domain~\citep{adlakha2022topiocqa}. In this paper, we have used prior research in intelligent systems (discussed in Section \ref{section-related-works}) to guide our research objectives and methodology. 
We use the Conversational Roles (COR) model to propose a pipeline connecting user utterances to system-level actions. We identify the speech acts for each utterance to identify user intent and consequently, use the speech acts to predict the system-level search actions. We perform ablation analysis to determine the important features for prediction and analyze the dataset for error analysis. To answer our research questions, we also conducted an extensive user study to collect user-system interaction data, and performed thematic analysis of the dialogues and search actions.

The rest of the paper is organized as follows: Section 2 discusses research goals and contributions, while Section 3 presents prior research work on related topics. Section 4 explains the data collection and annotation process while section 5 details our experimental methodology, from feature generation to model building. The results are presented in Section 6, along with the ablation analysis and statistical significance of our models. We also perform a detailed error analysis to account for the classification errors. Section 7 concludes the paper by stating limitations and future directions.

\section{Research Goals \& Contributions}

%key aspect of the paper
Our research aims to facilitate a better understanding of natural language in information-seeking scenarios and, subsequently, extend that understanding to intelligent search actions.
%major issues with understanding the conversation
When using natural language, users may not follow any guidelines. As such, the dialogues can be unclear, wordy, and unrestricted. 
%solution
Therefore, the agent should be able to handle vague, dynamic information needs and rapidly evolving search objectives~\citep{stein1995structuring}. 
A careful review of the prior literature revealed that %{\color{red}
limited research has been done to improve natural language understanding of conversational search systems. Almost no work has been done to find the connection between natural language understanding and system-level search actions. Therefore, the motivation of this research was to better understand the utterances (from a search system perspective), use the concepts of speech acts and search actions to guide system-level search actions, and validate the application of the deep neural model to smaller user study datasets (by augmenting the user-system dialogues with linguistic and metadata features).
%}

Before diving into our research questions, it is prudent to formally define two terms used in the paper -- Utterance and Speech Act (or Dialogue Act). 
Utterance has several definitions in communication theory and linguistics, but the definition that best represents our work was provided by \citet{belkin1987discourse}. According to ~\citet{belkin1987discourse}, \textit{an utterance is a sequence of speech which originates from one participant during a conversation. It may or may not contain complete grammatical entities and can be terminated by contribution of another participant. If the contribution of one participant takes the conversational turn, the previous speech sequence is regarded as completed, and a new utterance begins}~\citep{brooks1983using, price1983functional}.
Speech Act, also known as Dialogue Act or Illocutionary Act, was first introduced by~\citet{austin1962speech} and later researched by~\citet{searle1969speech} and~\citet{searle1979taxonomy}. \textit{The purpose of these acts is to classify human speech in order to predict the propositional content present in it.} Every utterance constitutes a certain action like praise, greeting, promise, criticism, threat, and so on. The speech acts were introduced to distinguish between utterances based on certain actions. 
Speech acts when used in the context of natural language conversations are also referred to as ``dialogue acts'' in the literature ~\citep{stolcke2000dialogue}. Since the terms have very little difference in meaning and functionality, we have used the two terms interchangeably in this paper.   

In line with our research objectives, we identified the following research questions: 
\begin{enumerate}[label=RQ\arabic*]
    \item \textbf{\textit{How can we improve the natural language understanding of user utterances in a conversational search setting 
    using speech acts}?}
    \item \textbf{\textit{ 
    After predicting the speech acts accurately, can we use them downstream to guide the search actions being performed by the agent?}}
\end{enumerate}

%goal
While designing an effective conversational search system, it is vital to correctly associate the utterances with different speech acts. By using the search intention associated with the speech acts, the system should be able to identify the information need of the user and the goal of the search. In other words, searching by talking is motivated by the functional aspects of speech. Therefore, in this research, our objective is to automatically classify the speech acts in a user-system information-seeking conversation and connect them to corresponding search actions.
% need to add a few lines about the paper and findings
To collect user-system information-seeking dialogues, we conducted an extensive user study involving 25 users and 75 search sessions. The user-system conversations -- containing 1,834 utterances and 509 system actions -- were analyzed for themes, and the speech acts were identified for each utterance. The search actions of the agent were also identified thematically. Our thematic analysis yielded a total of eight speech acts and four search actions. This paper presents the details of the user study, the dataset, and the thematic analysis. We also present the details of the features used and the automated classifiers developed to identify the speech acts and the corresponding system-level search actions. 
Our classification system is based on an attention-based deep neural network (ADNN), which combines three categories of features generated from the dialogue transcripts: word-level features using BERT, natural language features, and metadata. 

%For evaluation, 
To validate the generalizability of our approach, we annotated another publicly available dataset with respective speech acts and search actions.
Results of ablation analysis showed that the best-performing model (\textit{meta+linguistic+bert}) outperformed the single-category BERT model for speech act classification (on the CONVEX dataset). The model also performed reasonably well for search action prediction. We validated our findings using a second dataset (%{\color{red}
Spoken Conversational Search dataset or %}
SCS), which shows similar results. Our research should improve the system-level understanding of natural language and comprehend how spoken dialogues (by the user) trigger the search actions performed by the system.

There are two original contributions of this work in the field of conversational search systems:

\begin{enumerate}
    \item \textbf{\textit{Creation of gold standard conversational datasets}}: \\
    We develop two gold standard datasets with multi-turn conversations. Both datasets contain information-seeking conversations for non-factoid user queries. To create the first dataset, we collected multimodal data using a Wizard-of-Oz study and performed thematic analysis to annotate the utterances with speech acts and search actions. We annotated another publicly available dataset to validate our results.
    \item \textbf{\textit{Development of algorithmic models to connect user utterances to system-level search actions}}: \\
    We use the COR model -- a popular framework in discourse analysis and intelligent systems -- to propose a pipeline connecting user utterances to system-level actions. First, we identify the speech acts for each utterance to identify user intent. Next, we use speech acts to predict the system-level search actions. We perform ablation analysis to determine the important features for prediction and analyze the dataset to explain our results and errors.
\end{enumerate}

\section{Related Works}
\label{section-related-works}
In this section, we discuss the literature related to conversational search systems. 

\subsection{Defining Conversational Search Systems}

\citet{bunt1989information} is credited for coining the word {\em information dialogue} for the type of dialogues observed in simple information systems which provided factual information. When used in the context of IR systems, the term dialogue refers to the multiple rounds of negotiation or clarification that occur between the dialogue partners (user and intermediary). Such an interaction aims at developing a constructive solution to the initially vague information problem and hypothesizing the information need of the user~\citep{stein1999user}.
Although conversational search systems are often defined as artificial systems which can interact, understand, and respond in natural language~\citep{ram2018conversational, laranjo2018conversational}, it overlooks the complexity of a human-human conversation. Conversations are interactive and incremental, involving multiple rounds of turn-taking, explanatory and educational for both parties involved, and expeditious~\citep{joho2018cair}. 
\citet{radlinski2017theoretical} provides the first formal definition of conversational search systems.%:
To understand the intention and information needs of the user, both the user and the intermediary need to develop cognitive models of each other in an effective information transfer environment~\citep{radlinski2017theoretical, belkin1984cognitive}. To have successful communication, the participating individuals must possess a model of the other and negotiate it to perfection~\citep{be7dd91c763142bdb1c3a132e3a032ce}.

\subsection{Speech or Dialogue Act Classification}

Dialogue acts (or Speech acts) are used to understand the communicative intent of the user~\citep{saha2020towards}. The challenge of dialogue acts is to associate semantic labels with utterances for understanding the intention of the user~\citep{chen2018dialogue}. The composition of discourse is always context-sensitive, and therefore, the dialogue act of an utterance can always be inferred from the preceding utterance~\citep{lehrberger1982sublanguage}. 
Dialogue Act classification has been used widely for natural language understanding of a discourse~\citep{chen2018dialogue, bothe2018context, ghosh2021classifying}. In recent years, different machine learning algorithms and various approaches have been adopted to perform dialogue act classification for purposes like sentiment classification~\citep{saha2020towards}, slot filling~\citep{firdaus2021deep}, intent detection~\citep{firdaus2021deep}, and detecting sensitive dialogues in counseling sessions~\citep{malhotra2022speaker}. In earlier studies~\citep{reithinger1997dialogue, grau2004dialogue}, the focus was to classify dialogue acts based on lexical, syntactic, and prosodic features. Other works employed machine learning models like CRF~\citep{xue2018dialogue, li2019dual}, CNN~\citep{kim2018dialogue, wang2020dialogue}, and RNN~\citep{firdaus2021deep, yee2020myanmar}. Bidirectional-LSTMs -- among other RNN architectures -- were popular for such tasks because of its capability to retain the context of previous utterances. In recent studies, some researchers have used transfer learning from large transformer-based language models like BERT and RoBERTa~\citep{saha2020bert, wu2020tod}. Attention-based architecture has also been used in dialogue acts to perform slot filling and intent classification~\citep{firdaus2021deep}.

%%% COR MODEL -- FIGURE REMOVED FOR LACK OF SPACE 
\begin{figure}[!htpb]
  \includegraphics[scale=0.25]{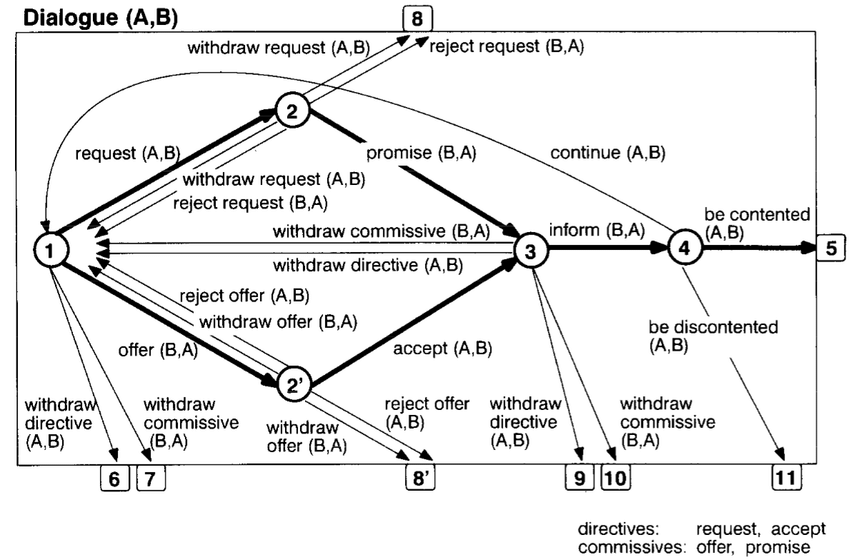}
  %COR.png}
  \caption{%\label{font-figure} 
  COR Model (Sitter and Stein, 1992).%}
  }
  \label{fig-COR}
\end{figure}
%REFERRED IN Section 3.3

\subsection{Theoretical frameworks}

Some early research explored how dialogues can be incorporated in retrieval systems~\citep{sitter1996modeling, stein1995structuring}. 
According to ~\citet{winograd1986understanding}, in an information-seeking episode, the searcher-intermediary dialogue is akin to a conversation and can be modeled as a state transition network which %This 
formed the basis of the Conversation for Action (CfA) model. 
Later, ~\citet{sitter1996modeling} introduced the application-independent COR model (Figure \ref{fig-COR}) 
-- for human-computer information-seeking scenarios. 
The COR model focuses on interpersonal dialogues and helps in the acquisition of knowledge about the conversational partners incrementally. 
The speech acts in the COR model were developed using prior work done by ~\citet{searle1979taxonomy,searle1985foundations}. 
In theory, any human-human search interaction can be mapped using the COR or the CfA model.
In a successful conversation, both the searcher and the intermediary need to collaborate and create models of each other incrementally~\citep{be7dd91c763142bdb1c3a132e3a032ce, stein1999user}. 

Some of the latest works in this domain attempted to develop frameworks capable of explaining the information-seeking dialogues and the associated cognitive functions~\citep{azzopardi2018conceptualizing, trippas2017conversational, vakulenko2017conversational} or for evaluation purposes~\cite{lipani2021doing}.
~\citet{trippas2017conversational} suggested a turn-based framework for the spoken environment, ~\citet{azzopardi2018conceptualizing} provided an extensive list of actions and interactions in conversational search while ~\citet{vakulenko2017conversational} proposed the QRFA model to show the conversation flow in information-seeking episodes. These frameworks created the platform for automatically identifying the different dialogue patterns and the corresponding search actions witnessed during a human-human information-seeking dialogue.

\subsection{Methodological Approaches}
%\textcolor{blue}{
Several studies have attempted to understand the user behavior and preferences for conversational agents~\citep{arguello2017factors, avula2018searchbots, yuan2011design, begany2015factors}. Several studies use crowdsourced workers~\cite{ferreira2022open} or Wizard of Oz techniques~\cite{thomas2017misc, trippas2017conversational, trippas2018informing}, and almost all of them monitor the interaction patterns between the searcher and the agent~\citep{vtyurina2017exploring, trippas2018informing, teevan2004perfect, dubiel2018investigating, ghosh2020exploring}. Some studies have also been conducted to evaluate and improve user satisfaction in conversational systems~\citep{kiseleva2017evaluating, kiseleva2016predicting, DBLP:journals/corr/abs-1709-04734, mehrotra2017hey, ghosh2021can}. 
Others have worked on querying by voice~\citep{utama2017voice}, reformulations of spoken queries~\citep{nogueira2017task} and their characteristics~\citep{guy2016searching}, and identifying user intent through query suggestions~\citep{radlinski2017theoretical}, clarifications~\citep{zou2023users, aliannejadi2021analysing, vakulenko2021large, salle2021studying, bi2021asking, aliannejadi2019asking}, useful question generation for leading a conversation~\citep{rosset2020leading} and negative user feedback~\citep{bi2019conversational}.

Machine (and deep neural) learning has been a popular choice for solving various problems in conversational IR~\citep{yan2018smarter, DBLP:journals/corr/abs-1709-04734, LiJiwei2017ALfN}. 
Deep neural networks have also been used to answer complex questions~\citep{DBLP:journals/corr/abs-1712-07229}, predict the success of dialogues~\citep{lykartsis2018prediction}, improve contextual awareness~\citep{yan2016learning}, reformulate multiturn questions~\citep{choi2018quac, gao2019neural}, topic propagation~\citep{mele2020topic}, present exploratory search results as interactive stories~\citep{vakulenko2017conversational} and result presentation of structured data over voice~\citep{zhang2020summarizing}. Recent works use transformers~\cite{ferreira2022open}, attentive networks~\cite{DBLP:journals/corr/abs-1712-07229}, and large language models~\cite{bi2021asking, ghosh2021classifying}.
A lot of work has also been done on conversational recommendation~\citep{sun2018conversational, li2018towards, micoulaud2016acceptability, zhang2018towards} where the authors use end-to-end frameworks for e-commerce, movie, music, healthcare, and banking industries. 
\citet{trippas2020towards} in their work does a qualitative analysis of user-agent interaction over audio channels. They propose incorporating interactivity and pro-activity to reduce the complexity of search tasks in spoken conversational search. The findings from this paper can be a reference framework for future conversational search systems.

\section{Data Collection and Dataset}

\begin{figure*}[!htpb]
    \centering
    %\includesvg[scale=0.4]{images/woz-1.svg}
    \includegraphics[scale=0.9]{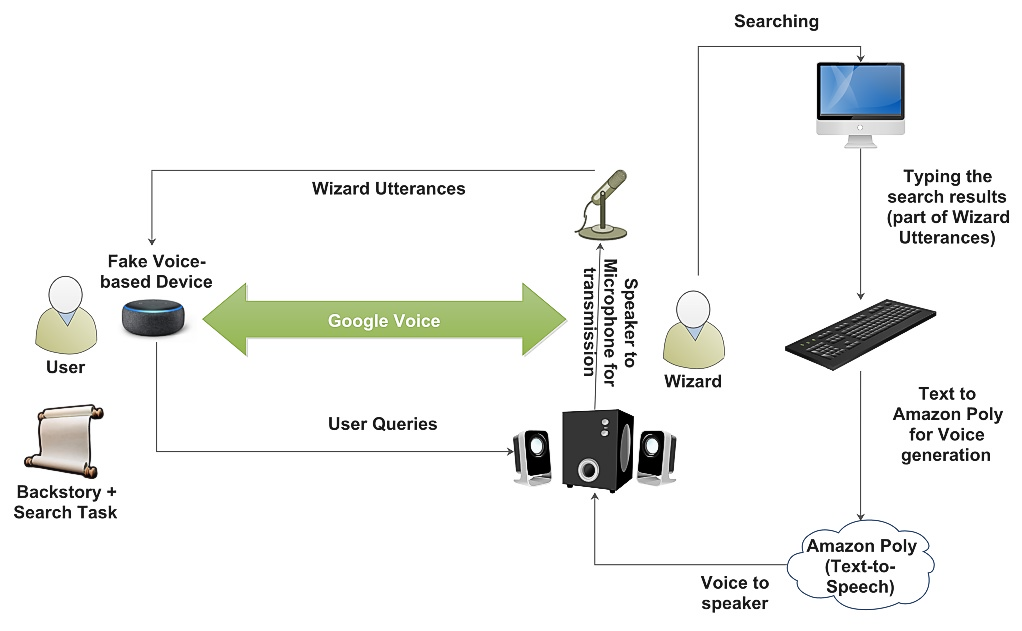}
    \caption{Experimental setup to conduct WOZ study}
    \label{WOZ setup}
\end{figure*}

%Our research question 
To achieve our research objectives, we required realistic user-system interaction data. While some of the publicly available datasets~\citep{trippas2018informing, thomas2017misc} contain conversational search dialogues, their research objectives did not necessitate hiding the human nature of the intermediary. Therefore, the collected data is more human-human than human-system and is unlikely to be replicated in a voice-based system in the near future.
The data collected as part of this study closely resembles a user-system interaction we might witness in a few years. The experimental setup that was used to conduct the Wizard-of-Oz study is shown in Figure~\ref{WOZ setup}.
In the following sections, we explain the details of the user study, the thematic analysis, and the annotation process performed to create the gold standard data.

%ACCESSED IN Section 4 before 4.1

\subsection{User Study}

To create the dataset, we conducted an in-laboratory user study with seventy-five information-seeking interactions between the users and the intermediary. 
%%new content
%{\color{red}{
To recruit participants, we circulated a call for recruitment at libraries and campus dining halls, electronically over university emailing lists, and on online forums. Although previous studies \citep{white2006implicit} have shown that experienced and inexperienced users perform search differently, we preferred participants who were fluent speakers and listeners of North American English, experienced in using the internet and search functions, and familiar with voice-based conversational search systems. The preferred language skills and search expertise of the participants were advertised in the call -- and verified through self-reporting -- but not assessed by the researchers. 
The role of the intermediary was most vital for the success of our WOZ study. The Wizard needed to be an experienced searcher who performed searches and provided results in real-time but also ``a con man" \citep{price2002off} to deceive the user into thinking that the intermediary is non-human. A computer science undergraduate student with proficiency in searching online and experience in performing voice searches was selected for the role of Wizard.  

Twenty females and six males, including the Wizard, were recruited for our study. The maximum and minimum ages of our participants were 29 and 19 years, respectively, with a median age of 21, a variance of 8.15, and a standard deviation of 2.855. Twenty-two individuals described themselves as native English speakers, while the remaining three listed their original languages as Greek, Hindi, and Gujarati, respectively. The participants self-reported their skills in speaking and hearing English as proficient (with means of 4.8 and 4.92, respectively, on a 5-point Likert scale ranging from 1 (novice) to 5 (expert)). The average online web search skill was 4.6, and the median was 5. Almost all participants had prior experience with voice-based personal assistants and rated their success rate between 1 and 5 (with a mean of 3.2 and a median of 3.0).

%%Task development
The study involved three search tasks and twenty-five participants (users). 
% REORDERED FOR CLARITY
%Two search tasks were moderately complex (Conference and Perfume search), while the third task (Sickness) was cognitively simple. The tasks were rotated to avoid learning effects.
We developed the tasks following prior literature \citep{white2006implicit, kelly2015development}, and tasks were designed to simulate naturalistic search behavior among the users. There were three tasks -- one warm-up task at the beginning and two main tasks which were rotated to avoid task learning effects. All the tasks were developed following Bloom's Taxonomy. The warm-up task belonged to the Remember and Understand levels of Bloom's Taxonomy~\citep{bloom1956taxonomy} (Health Information Search), while the main tasks (Conference Search and Perfume Search) belonged to the Analyze and Evaluate levels. The tasks were adopted from~\citet{white2006implicit} and~\citet{kelly2015development}, and they initiated a multi-turn conversation between the user and the intermediary. The health information search required the user to identify facts, while the experimental tasks required them to compare options and make a recommendation (for more details about tasks, see~\citet{ghosh2020exploring}). While present conversational systems are more adept at handling factoid questions, it was essential to develop search tasks that would be more complicated and represent search tasks for future systems.

Our task design required us to reduce variances caused by multiple users operating as Wizards. Therefore, we employed a single user -- experienced in searching online and using conversational systems -- to play the role of the Wizard. 
To eliminate any influence of task and topic learning effects, and personal attributes (of the Wizard), we developed a set of protocols for the Wizard~\cite{ghosh2021can}. The protocols -- along with training -- ensured that the search behavior and skill of the Wizard did not change during the course of the study. Before conducting the user study, the Wizard was provided with the task descriptions and was allowed to search online and prepare notes about the topic. The training made the Wizard more familiar with the experimental setup and improved his task and topic knowledge. It also allowed him to consider the possible search directions and responses. In addition, a script -- containing predefined templates of dialogues for different search situations -- was also provided to the Wizard to standardize the vocabulary. These templates replicated the simple vocabulary commonly used by artificial conversational systems. For example, every time the user started a search session, the Wizard responded with ``Hi, I am Joanna; how may I help you today?''. Similarly, search sessions ended with ``It is always great talking to you, bye!''

The actions of the Wizard were restricted to maintain the pretense: \textit{the user thought that he was interacting with an intelligent automated agent who is smarter than the existing state-of-the-art but not as smart as a human replacement.} Our approach ensured that we captured an authentic human-computer interaction without developing the actual state-of-the-art system. 
The user and the Wizard were located in separate rooms. The user had access to the mock system, while the Wizard had access to a networked computer. The audio channel between the user and the Wizard was established using Google Voice. 
%\footnote{\url{https://voice.google.com/}}. 
Google Voice also allowed us to record the entire conversation between the user and the Wizard.
The Wizard searched on the computer and typed in the response in textual form. The text was then converted to speech using Amazon Polly 
%\footnote{\url{https://aws.amazon.com/polly/}} 
and played back to the user. The search screen of the Wizard was captured using Kaltura.% \footnote{https://www.kaltura.org/}.

\subsection{Thematic Analysis and Annotation}

Each search session ranged from 5 to 20 minutes, with a total of around 10 hours of audio and video to transcribe.
Our first step was synchronizing the audio (user-system dialogues) with the video (search actions by the Wizard) using an open-source video editing tool. 
We uploaded the processed audio file obtained from the last step to Amazon Cloud %\footnote{https://aws.amazon.com/s3/}
and used Amazon Transcribe %\footnote{https://aws.amazon.com/transcribe/}
to transcribe the audio files automatically. Automatic transcription converted user and agent utterances from speech to text, generated timestamps for each utterance, and labeled the speakers. All automatically generated data was verified to ensure that the integrity of the data was not compromised during the automated transcription process. For transcription, we followed the steps highlighted in previous works~\citep{mclellan2003beyond, thomas2017misc, trippas2017conversational} and made necessary changes as required based on our data. 
The user and system dialogues were split or merged manually based on the definition of utterances in literature~\citep{brooks1983using, price1983functional}. Lastly, the video captured %using Kaltura
was observed, and the search actions performed by the Wizard were added to the dataset with the corresponding timestamps. 

The researchers performed qualitative coding of the data to identify the underlying themes using steps proposed by ~\citet{braun2006using}. The analysis was performed on the data we collected and on another publicly available dataset~\citep{trippas2018informing,trippas2017conversational}%,trippas2017people}
, which we used to validate our classification model. The initial speech and search actions were annotated using frameworks proposed in~\citet{trippas2017how} and ~\citet{azzopardi2018conceptualizing}, and were refined subsequently since the data collected and the themes proposed by~\citet{trippas2020towards} are more suitable for conversations where the user is aware of the human nature of the intermediary. Our final set of themes contained 12 speech acts and 4 search actions. 
There were three annotators, and the inter-annotator agreement (Cohen's Kappa) was 0.861 for speech acts and 0.932 for search actions (Refer to Table~\ref{tab-kappa} for details).

% \usepackage{booktabs}
\begin{table}[]
\centering
\caption{%{\color{red}
Inter-Annotator Agreement}
%}
\label{tab-kappa}
\begin{tabular}{@{}llllll|ll@{}}
\toprule
\multicolumn{6}{c|}{Speech   Acts} & \multicolumn{2}{c}{Search Actions} \\ \midrule
Code & \multicolumn{1}{l|}{$\kappa$} & Code & \multicolumn{1}{l|}{$\kappa$} & Code & $\kappa$ & Code & $\kappa$ \\ \midrule
S1 & \multicolumn{1}{l|}{0.93} & S5 & \multicolumn{1}{l|}{0.81} & S9 & 0.83 & SR1 & 0.98 \\
S2 & \multicolumn{1}{l|}{0.96} & S6 & \multicolumn{1}{l|}{0.69} & S10 & 0.92 & SR2 & 0.93 \\
S3 & \multicolumn{1}{l|}{0.40} & S7 & \multicolumn{1}{l|}{0.67} & S11 & 0.39 & SR3 & 0.94 \\
S4 & \multicolumn{1}{l|}{0.94} & S8 & \multicolumn{1}{l|}{0.80} & S12 & 0.93 & SR4 & 0.90 \\ \bottomrule
\end{tabular}
\end{table}
%Table referred in section 4.2

\subsection{Themes for Speech Acts and Search Actions}
\label{section-themes-speech-and-search-acts}

\begin{figure}[!htpb]
	\centering
	\begin{subfigure}[t]{0.49\linewidth}
		\centering
        \includegraphics[scale=0.33]{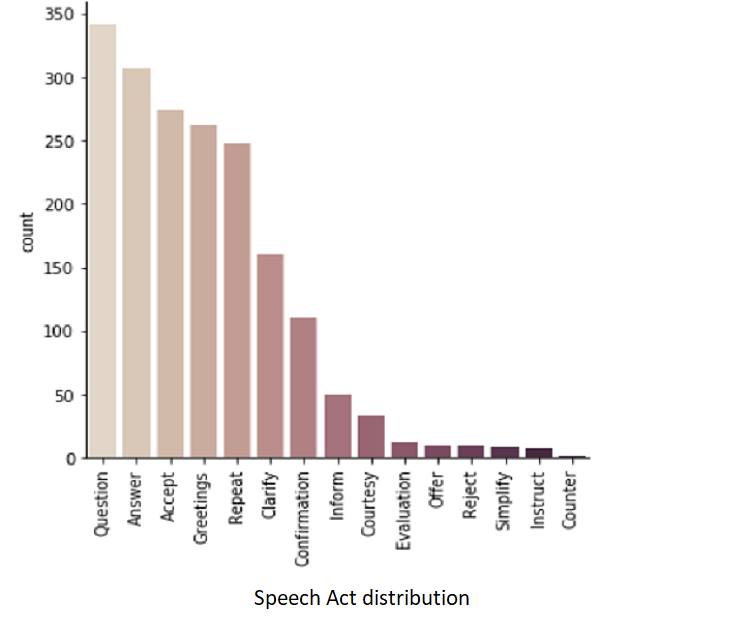}
        \caption{Speech Acts (CONVEX).}
        \label{fig:2a}
	\end{subfigure}%
	~
	\centering
    \begin{subfigure}[t]{0.49\linewidth}
		\centering
        \includegraphics[scale=0.33]{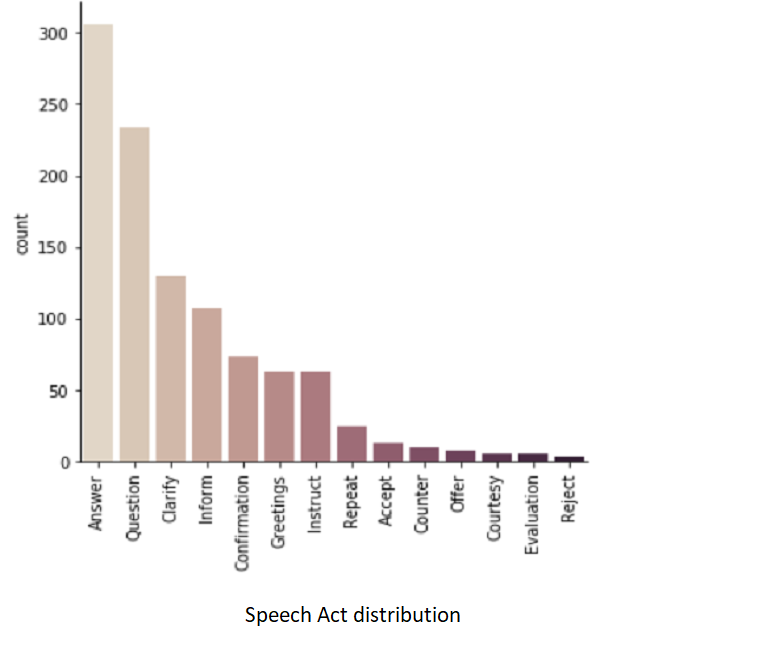}
        %[width=\linewidth, height = 0.12\paperheight]
        \caption{Speech Acts (SCS).}
        \label{fig:2c}
	\end{subfigure}%
    
	\caption{Speech Acts for CONVEX \& SCS Datasets.}
    \label{fig-dataset-statistics-speech}
\end{figure}

\begin{figure}[!htpb]
    \begin{subfigure}[t]{0.49\linewidth}
		\centering
        \includegraphics[scale=0.33]{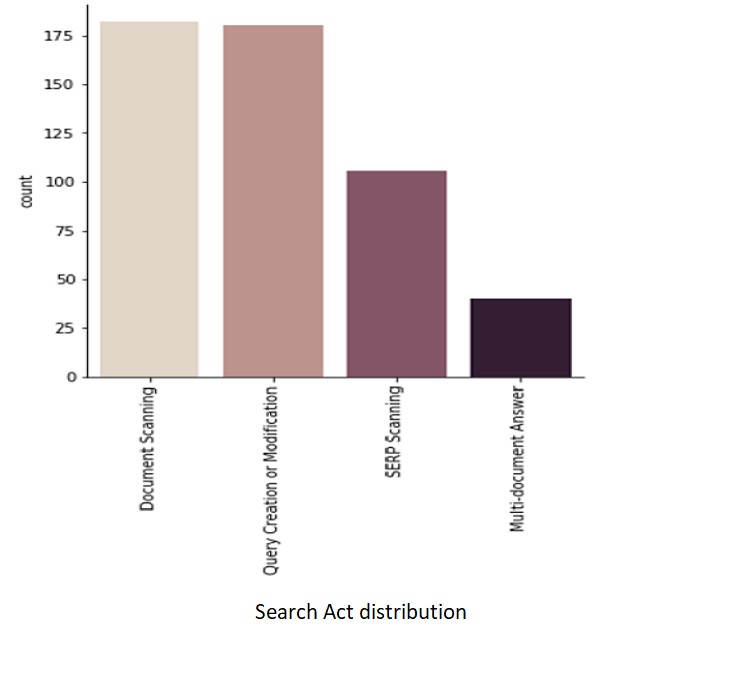}
        %[width=\linewidth, height = 0.12\paperheight]
        \caption{Search Actions (CONVEX).}
        \label{fig:2b}
	\end{subfigure}%
	~
    \begin{subfigure}[t]{0.49\linewidth}
		\centering
        \includegraphics[scale=0.33]{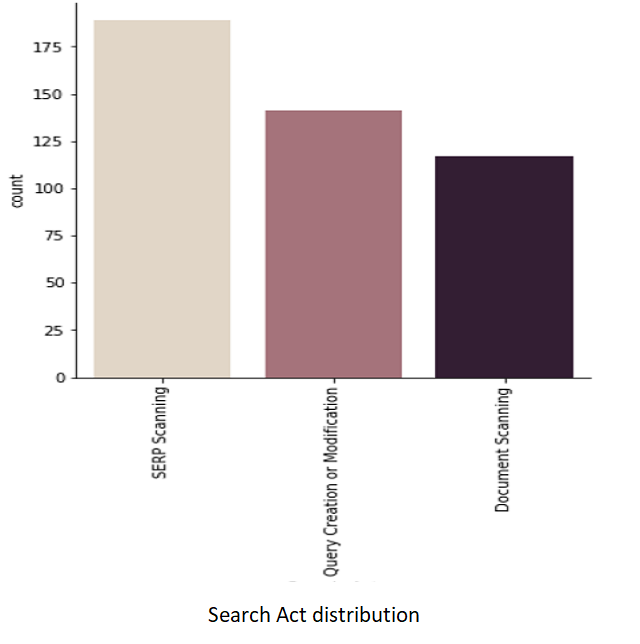}
        \caption{Search Actions (SCS).}
        \label{fig:2d}
	\end{subfigure}%
	
	\caption{Search Actions for CONVEX \& SCS Datasets.}
    \label{fig-dataset-statistics-search}
\end{figure}

We initially identified 14 categories for Speech Acts %(Table~\ref{tab:initial-speech-acts}) 
and 9 categories for Search Actions based on prior literature~\citep{stein1999user,winograd1986understanding, trippas2017how,azzopardi2018conceptualizing}. 
After two rounds of thematic analysis, the researchers finalized the themes for Speech and Search Acts. The final dataset contains 12 speech acts (see Section 4.3.1) and four search actions (see Section 4.3.2). The frequency of the identified speech acts and search act in the CONVEX dataset and the SCS dataset are presented in Figure~\ref{fig-dataset-statistics-speech} and Figure~\ref{fig-dataset-statistics-search} respectively.

\subsubsection{Speech Acts}

The description of the Speech Act themes we identified for the final annotation scheme is described below:
\begin{enumerate}
    \item Question or Seek: The utterance contains the initial information request. It could also involve the situation when the user comes up with a new search request during the conversation. 
    Example: \textit{``Joanna I am looking for men's perfume can you give me some example?.''}
    
    \item Accept or Reject: The utterance is used to accept or reject the request of the conversational partner. 
    Examples: \textit{``Ok, please let me look into it. Give me a few minutes.''} and \textit{``I will not be able to answer that.''}

    \item Counter or Offer: Once the agent develops a better understanding of the user's information problem, the agent may suggest a modification to the query or offer to do something different from the user's request. It could also be a request to simplify the search query if the original query is too complex for the agent. The control of the conversation transfers from the user to the agent when this speech act is encountered.
    Examples: \textit{``You would have to name a specific conference so I can check the deadline.''}, \textit{``None of these are in Europe. Would you like me to query Top Conferences in AI early 2020 in Europe?''}, or \textit{``Hmmm. That search has become too complex for me. Can we do it in steps?''}
    
    \item Answer: The agent either informs the user of the search result or answers the question asked. This act signals the transfer of control back to the user. It could either be the final answer to the user's problem or an intermediate step. 
    Example:\textit{ ``Yes, the \$8.95 shipping fees for the 2 to 3 day express shipping.''} or \textit{``According to the sephora.com, Yves ST Laurent l'Homme Cologne Bleue is 116 U. S. Dollars and contains Bergamot, Marine accord and Cardamom scent.''}
    
    \item Clarify: The agent seeks clarification from the user to better understand the user's information needs. It could either be an explicit request for confirmation or a follow-up question.
    Example: \textit{``Can you tell me more about what you are looking for?''}
    
    \item Inform or Declare: User utterances that provide additional information related to the search either clarify the question being asked or voluntarily add context to the information problem. 
    Example:\textit{``No, I want to know the price in US dollars.''}

    \item Evaluation: These user utterances suggest that the user is either content with the answer provided by the agent or unsatisfied with the results. 
    Example: \textit{``That\'s the one.''}
    
    \item Instruct: A direct instruction from the user to the agent on how to perform the search. This could be done by defining keywords, queries, information sources, or search strategies. At this point, the control of the search is with the user instead of the agent.
    Example: \textit{``No. Query, how many artificial intelligence conferences is [sic] there are in the United States?''}
    
    \item Repeat: The agent may ask the user to restate the information request, or the user may ask the agent to repeat the last utterance or answer. 
    A common occurrence of this speech act was observed when the answer provided by the agent was too long and the user could not follow. Similarly, the agent requested the user to repeat when the user's utterance was unclear or verbose. 
    Example:\textit{``Can you repeat that?''}
    
    \item Confirmation: 
    These user utterance confirms -- either positively or negatively -- when the agent asks for clarification or feedback.
    Example: \textit{``Yes. that is ok.''} and \textit{``No, I don't.''}

    \item Courtesy: The user or the agent follows the norms of polite conversation by being deferential. 
    Example: \textit{``Is there anything else I can help you with?''}
    
    \item Greetings and Closing Rituals: 
    These utterances contain key phrases spoken by the user to activate and end the search session. They may also include greetings uttered by the agent at the beginning and the end of each search session. 
    Example: \textit{``Hi Joanna.''} and \textit{``It is always great talking. Bye.''}

\end{enumerate}

%REFERRED to in Section 4.3 before 4.3.1

\subsubsection{Search Actions}

The top-level themes finalized for search actions are listed as follows:
\begin{enumerate}
    \item Query Creation or Refinement: 
    The agent creates a new query or modifies an existing query for subsequent search. 
    Example: \textit{Google search with the term ``bergamot and lavender cologne''.}
    
    \item SERP Scanning: 
    The agent scans the search engine results page (SERP) and summarizes the top results. It may include a summary or answers provided by the search engine at the top of the results page. 
    Example: \textit{Google's rich answer to the query: ``238 pounds to USD'':  Tom Ford Private Blend Venetian Bergamot is 306.52 United States dollars.}
    
    \item Document Scanning: The agent reads from inside the documents returned by the current query or from a previously opened document.
    Example: \textit{[Reading from sephora.com] ``According to the sephora.com Yves ST Laurent l'Homme Cologne bleue is 116 U. S. Dollars and contains Bergamot, Marine accord and Cardamom scent.''}
    
    \item Organizing Answers from Multiple Documents: The agent reads and summarizes answers from multiple documents which were returned by the query.  
    Example: \textit{[Combining answers from cigna.com and healthline.com] ``There is no permanent cure for allergic rhinitis. One of the best things you can do is to avoid the things that cause your allergies. You can take antihistamines to treat allergies. You can also use decongestants to relieve a stuffy nose and sinus pressure. Eye drops and nasal sprays can help relieve itchiness. Your doctor may recommend immunotherapy, or allergy shots if you have severe allergies.''}

\end{enumerate}

%REFERRED to in Section 4.3 before 4.3.1

\subsection{CONVEX Dataset}

\begin{figure}[!htpb]
	\centering
    \includegraphics[width=0.7\linewidth]{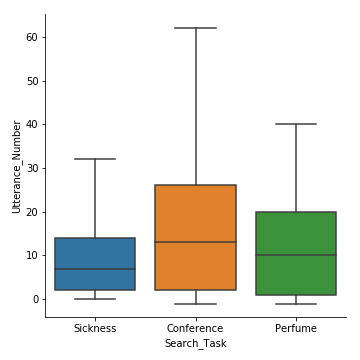}   
    \caption{Frequency of utterances for search tasks.}
    \label{fig: convex-utterances-by-tasks}
\end{figure}

The final dataset (CONversation with EXplanations or CONVEX)\footnote{https://github.com/SouvickG/CONVEX} comprised 768 user utterances and 1066 intermediary utterances, each representing a speech action. The utterances are interspaced with 509 search actions performed by the intermediary. 
Figure~\ref{fig: convex-utterances-by-tasks} compares the number of utterances for each search task.
The CONVEX dataset contains 75 search sessions with 1,834 speech acts and 509 search actions. While a bigger sample size is always desirable, the size of our data allows us to perform statistical analysis with the requisite power. We have also used transfer learning on a pre-trained DistilBERT model (more details in Section 5) that requires much fewer training instances~\citep{mosbach2021stability} than untrained BERT and helps us avoid any detrimental impact of the smaller dataset on our classification system. 

Finally, to evaluate generalizability, we have validated our results against another publicly available dataset~\citep{trippas2018informing, trippas2017conversational, trippas2017how} (Spoken Conversational Search or SCS) \citep{trippas2017conversational}. SCS contains 1044 speech actions, of which 527 are user utterances, and 516 are intermediary utterances. The intermediary performs 447 search actions.

The speech acts apply to both the user and the agent, while the search actions were performed only by the agent. 
In addition to the utterances and the speaker information, the dataset contains the user number, the search task topic, the task order, the timestamp and duration of each utterance, and the annotated themes for speech act and search action (where applicable). 

%REFERRED in Sectopm 4.4 first para

\section{Experimental Methodology}

This section discusses the various categories of features used, the architecture of the attention-based network, and the details of the hyper-parameters used.

\subsection{Feature Categories}

We have used three different feature categories which are used individually as well as in every possible combination through an attention-based deep neural network. 
Each feature category -- BERT, Linguistic features, and Metadata -- represents a different aspect of the data.

\subsubsection{BERT:}

Transformer-based machine learning architecture~\citep{vaswani2017attention} is fast as it facilitates parallel training and can capture long-range sequence features~\citep{wolf2020transformers}
, and has, therefore, emerged as a preferred choice for many Natural Language Processing tasks~\citep{wolf2020transformers} (like Machine Translation, Natural Language Understanding, and Natural Language Generation). 
BERT (Bidirectional Encoder Representation from Transformers), which is an application of the Transformer architecture, was open-sourced by Google in 2018~\citep{devlin2019bert}. Previous work has highlighted that BERT can overcome the shortage of training data by utilizing enormous amounts of unlabelled data available online~\citep{mosbach2021stability}.
Language models often benefit from pre-training~\citep{radford2018improving, dai2015semi} and can be used for a specific task with little fine-tuning. BERT exploits this attribute of language models and is pre-trained on BookCorpus (800M words)~\citep{zhu2015aligning} and English Wikipedia (2500M words). 
BERT also uses bidirectional self-attention and is capable of attending to every token and the context to its left and right. Single words cannot represent the meaning of a sentence individually; hence a model must understand how the words relate to each other in the context of a sentence. The attention mechanism helps BERT to make the composite representation of sentences rather than singular words. BERT is currently considered a state-of-the-art architecture as it has achieved a GLUE score of 80.5\%, MultiNLI score of 86.7\%, and SQuAD v1.1 and SQuAD v2.0 Test F1-scores of 93.2\% and 83.1\% respectively~\citep{devlin2019bert}.

%question-answering Test F1-score of 93.2, and SQuAD v2.0 Test F1 of 83.1\% 

In our experiment, we have used a distilled version of BERT called DistilBERT~\citep{sanh2019distilbert}, which is 40\% smaller than the actual BERT-base while achieving 97\% of its accuracy (as on GLUE benchmark). The speed of DistilBERT (60\% faster than BERT-base) was also a major consideration for choosing it over the BERT-base. DistilBERT is pre-trained on Toronto Book Corpus and full English Wikipedia. For Speech Act classification, it performs reasonably well when used as an isolated channel. However, for search action classification, the performance deteriorates when using BERT alone.

\subsubsection{Linguistic Features:}

For each utterance, the linguistic features were generated using the SpaCy API.%\footnote{https://spacy.io/}. SpaCy has been previously used for feature extraction in research related to natural language understanding~\citep{bocklisch2017rasa} and chatbots~\citep{ALTINOK18.3,khatri2018advancing}.
It is an industrial-strength API that is fast, effective, and performs reasonably well for NLP tasks.
The following linguistic features were generated from the user- and system- utterances at the token-level:

\begin{enumerate}
    \item The location and type of named entity;
    \item The type of character present in the token (for example, the alphabet, digit, punctuation, URL, stop-word, or out of vocabulary);
    \item The fine-grained and coarse-grained parts-of-speech for each token;
    \item The syntactic dependency relations of each token; and
    \item The categorical distance of the token from the beginning of the sentence.
\end{enumerate}

We generated word- or token-level one-hot representations of the above features, which were combined hierarchically to obtain a sentence-level representation of features, and then combined further to get utterance-level representations.

\subsubsection{Metadata:}

Metadata are the additional information available for data which helps to describe it better. In our experiment, several dialogue metadata were available. Most of these metadata are available for any user-system dialogue. Some metadata features could have been specific to our dataset, but it should not be challenging to derive them for any conversational data. The following metadata features were generated for our data: 
\begin{enumerate}
    \item Utterance Number: In any conversation, utterances are generated sequentially as the user and the system take turns in the conversation. The utterance number signifies the sequence of the utterance in the conversation. 
    
    \item Duration of utterance: The difference between an utterance's start and end times. 
    
    \item Speaker Role: The speaker could be the searcher or the intermediary.
    
    \item System: It refers to the system which was being used for the task. This is specific to the study; for example, CONVEX used two systems while SCS used only one. 
    
    \item Task complexity: It is the complexity of the search task. Once again, this was specific to the study. For example, CONVEX had search tasks of two complexity levels: 0 (low) and 1 (moderate). 
    In the SCS dataset, the task complexities were 1 (Remember), 2 (Understand), and 3 (Analyze). 
    
    \item Previous Speech Act: It is the speech act of the previous utterance.
    
    \item Previous Search Action: It is the last search action performed by the intermediary.
    
    \item Previous User Speech Act: The speech act of the last user utterance.
\end{enumerate}

\subsection{ADNN Architecture for Classification}

Attention layers are mainly of two types: additive~\citep{bahdanau2014neural} and scaled dot-product~\citep{vaswani2017attention}. In this study, we have used an additive attention mechanism to build our models for speech act and search action classification. Attention mechanism has been used in various fields like machine translation~\citep{bahdanau2014neural}, language inference~\citep{rocktaschel2016reasoning}, question-answering system~\citep{chen2017reading}, and document classification~\citep{yang2016hierarchical}.

\begin{figure*}[!htpb]
    \centering
    %\includesvg[scale=0.33]{images/Speech-to-Search.svg}
    \includegraphics[scale=0.81]{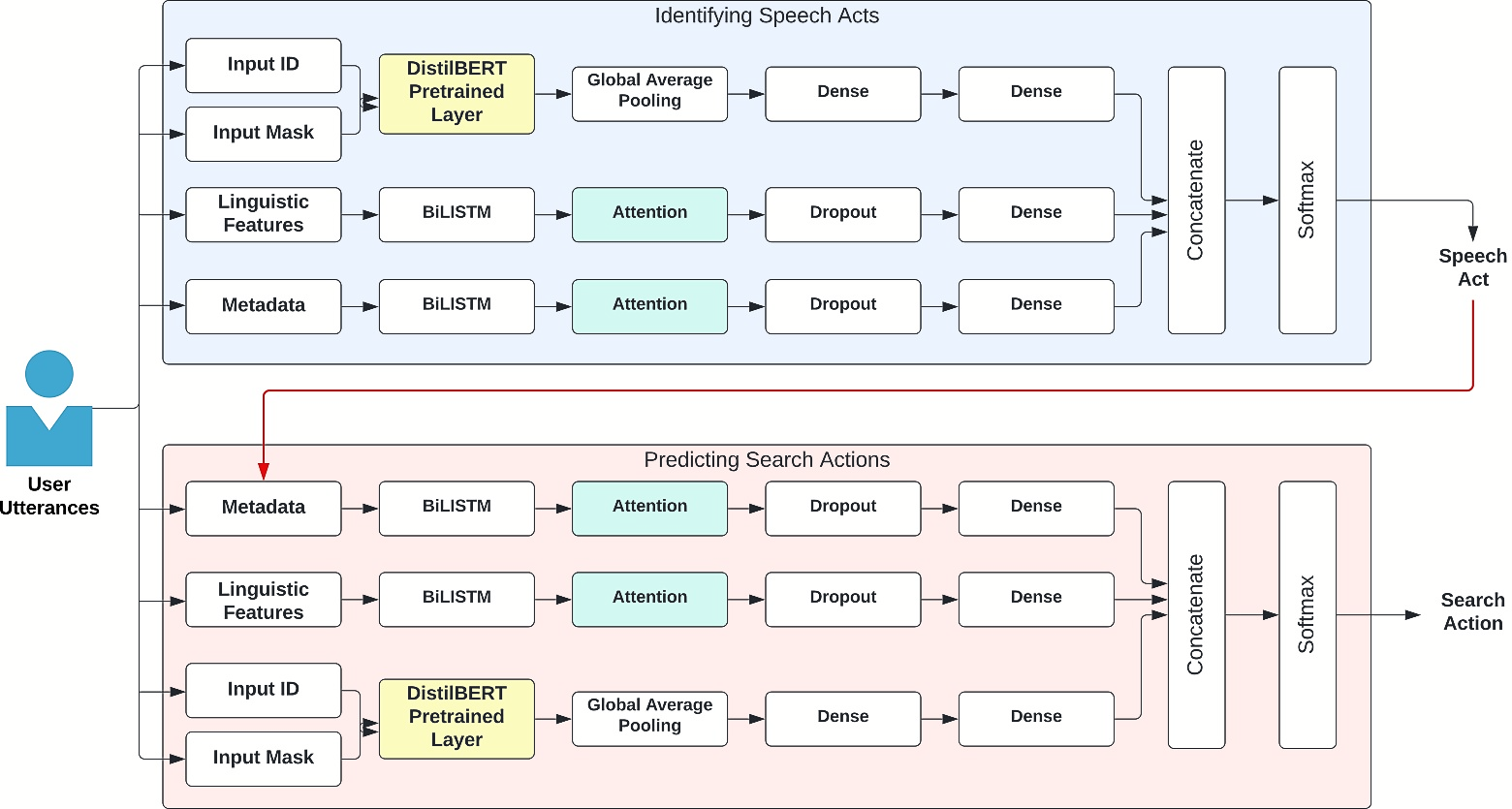}   
    \caption{Model Architecture with Three Feature channels}
    \label{fig:model_archi}
\end{figure*}
% REFERENCED in end of Section 5.2 just before Section 6

In ADNN (Attention-based Deep Neural Network), we have used Bidirectional LSTM (BiLSTM) in combination with the attention mechanism. The attention mechanism attends specific features more than the others, that is, attention weights~\citep{bahdanau2014neural} are calculated for every element in the input vector to find which subsets of features are more important than the others. Regular LSTM cells are unidirectional feed-forward networks that can process the input vector from left to right~\citep{schuster1997bidirectional}, whereas BiLSTM can analyze the vector from both directions. The bi-directional approach increases the number of parameters of the network, which allows the network to understand the sequences better and recognize patterns more efficiently. Our BiLSTM layer had a dropout of 0.25 and a recurrent dropout of 0.1. After weight modification by the attention layer, we used another dropout of 0.25 and a dense output layer with softmax activation (with an output dimension of n, where n depended on the prediction task). %The BiLSTM layer has a sigmoid activation function. 
The mathematical expression for calculating the attention weight is similar to the additive attention layer introduced by~\citet{bahdanau2014neural}. We used Adam as the optimization algorithm as it uses mini-batches to adjust the learning rates of model parameters. 
It is also more robust in choosing hyperparameters and considers estimates of the first and second-order moments for bias-correction~\citep{goodfellow2016deep_new}. 
Our model used categorical cross-entropy as a loss function to predict multiple classes (12 for speech acts and 4 for search actions). 
The ADNN architecture was used for linguistic features and metadata, while the pre-trained DistilBert~\citep{sanh2019distilbert} was fine-tuned to capture the word-level features in the dialogues. 
A simple representation of the model with the combination of all the three feature categories (\textit{meta+linguistic+bert} model) is shown in Figure~\ref{fig:model_archi}.

\section{Result and Discussion}

%Each model was trained and tested 30 times and we reshuffled the data every time. The tables provided in this section lists the different statistical measures of accuracy for each model.

In this section, we discuss the performance of our model in speech act and search action classifications for both the CONVEX and the SCS datasets. We also performed ablation analyses to study the importance of each set of features, which gives us 
an in-depth understanding of the features most useful for natural language understanding. 
We repeated the experiment thirty times, where the training and test sets were selected using thirty randomly generated seed values.

\begin{table*}[b]
\centering
\caption{Accuracy for Speech Act Classification}
%on CONVEX \& SCS dataset
\label{tab:convex_scs_speech}
\begin{tabular}{p{3.75cm}p{0.8cm}p{0.8cm}p{0.8cm}p{0.8cm}}%{@{}lcccc@{}}
\toprule
 & \multicolumn{2}{c}{CONVEX}  & \multicolumn{2}{c}{SCS} \\ \midrule
\textit{Feature Categories}& \multicolumn{1}{l}{\textbf{Maximum}} & \multicolumn{1}{l}{\textbf{Median}} & \multicolumn{1}{l}{\textbf{Maximum}} & \multicolumn{1}{l}{\textbf{Median}} \\ \midrule
\textit{\textbf{bert}} %\textsuperscript{*}
& 0.869& 0.822   & 0.602& 0.544   \\
\textit{\textbf{linguistic}} %\textsuperscript{\#} 
& 0.831& 0.786   & 0.665& 0.567   \\
\textit{\textbf{meta}}\textsuperscript{\ }  & 0.684& 0.644   & 0.564& 0.487   \\
\textit{\textbf{linguistic+bert}}  & 0.872& 0.832   & 0.584& 0.501   \\
\textit{\textbf{meta+bert}} %\textsuperscript{\#} 
& 0.913& 0.841   & 0.660& 0.500   \\
\textit{\textbf{meta+linguistic}}\textsuperscript{\ }  & 0.896& 0.853   & 0.689& 0.627   \\
\textit{\textbf{meta+linguistic+bert}}\textsuperscript{\#} & 0.902& 0.865   & 0.727& 0.635   \\ \midrule
\multicolumn{5}{l}{\begin{tabular}[c]{@{}l@{}} \textsuperscript{\#} and  \textsuperscript{\$} shows  significance at p\textless{}0.05 for CONVEX and SCS respectively \\ 
%\textsuperscript{\$} significance at p\textless{}0.05 for both
\end{tabular}} \\ \bottomrule
\end{tabular}
\end{table*}
%REFERENCED IN Section 6.1 last para

\subsection{Speech Act Classification}

The accuracy achieved by using various feature %channels
categories 
for speech act classification (on CONVEX data) is given in Table \ref{tab:convex_scs_speech}. Two models have achieved high accuracy: \textit{meta+linguistic+bert} and \textit{meta+linguistic}. The mean and median values of accuracy for \textit{meta+linguistic} model are 85.4\% and 85.3\%, respectively. The \textit{meta+linguistic+bert} model has slightly better average performance than the former, with mean and median accuracy values of 86.3\% and 86.5\%, respectively. %, 0.9 points, and 1.2 points of absolute difference. 
The ablation analysis shows that combining three feature categories yields the best result for speech action classification in CONVEX. 
We used a non-parametric Wilcoxon signed-rank test for pairwise significance assessments.

On deeper inspection and analysis of our dataset, we found some interesting findings related to the misclassified utterances. For example, four instances of S9 (Repeat) were classified as S1 (Question/Seek). The user requested the agent to repeat, which is very similar to a question, for example: ``Can you please repeat that?''. Similarly, many instances of S4 (Answer) were wrongly classified as S9 (Repeat). Our model could not differentiate between the system's answer to the user and the repetition of that answer when requested. Human annotators labeled the answers by the agent as S4, and any further repetitions %by the agent 
of the same answer as S9. For two utterances, the model marked S11 (courtesy) as S2 (accept), and these utterances only contained the word ``Okay.'' In three cases, clarification (S5) was confused as a question (S1). An example of a wrongly classified utterance is: ``It's a perfume for men or women?''; meanwhile, a correctly classified clarification utterance is: ``Based on what you said, I am running the query top conferences for artificial intelligence. Is that okay?''.

We used the same network architecture and verified our model on a second dataset (SCS) to verify how the model would perform for conversational searches under different experimental conditions. 
The results (Table \ref{tab:convex_scs_speech}) highlight that the highest accuracy was obtained using a combination of all the three channels (\textit{meta + linguistic + bert}), followed by \textit{meta + linguistic} and \textit{meta + bert}.  
\textit{Meta + linguistic + bert} model achieved a highest accuracy of 72.7\%  and a median accuracy of 63.5\% while those for \textit{meta + linguistic} were 68.9\% and 62.7\% respectively.  
The results reaffirm our belief that combining the three sets of features should provide the best model for understanding user dialogues, 
and no single-channel model could outperform the result achieved by concatenating the feature channels.
None of the models showed significant improvement for SCS.

\subsection{Search Action Classification}

\begin{table*}[!htpb]
\centering
\caption{Accuracy for Search Act Classification}
\label{tab:convex-scs-search}
\begin{tabular}{@{}lcccc@{}}
\toprule
   & \multicolumn{2}{c}{CONVEX}   & \multicolumn{2}{c}{SCS}  \\ \midrule
\textit{Feature Channel}   & \multicolumn{1}{l}{\textbf{Maximum}}& \multicolumn{1}{l}{\textbf{Median}}& \multicolumn{1}{l}{\textbf{Maximum}}& \multicolumn{1}{l}{\textbf{Median}}\\ \midrule
\textit{\textbf{bert}} & 0.382   & 0.294  & 0.511   & 0.422  \\
\textit{\textbf{linguistic}}   & 0.353   & 0.314  & 0.533   & 0.400  \\
\textit{\textbf{meta}}\textsuperscript{\$} & 0.588   & 0.51   & 0.611   & 0.544  \\
\textit{\textbf{linguistic+bert}}  & 0.422   & 0.328  & 0.511   & 0.428  \\
\textit{\textbf{meta+bert}}& 0.48& 0.358  & 0.533   & 0.456  \\
\textit{\textbf{meta+linguistic}}\textsuperscript{\ }  & 0.588   & 0.525  & 0.544   & 0.456  \\
\textit{\textbf{meta+linguistic+bert}}\textsuperscript{\#} & 0.588   & 0.515  & 0.533   & 0.444  \\ \midrule
\multicolumn{5}{l}{\begin{tabular}[c]{@{}l@{}}\textsuperscript{\#} and  \textsuperscript{\$} shows  significance at p\textless{}0.05 for CONVEX and SCS respectively \\ 
%\textsuperscript{\$} signifies significance at p\textless{}0.05 for both
\end{tabular}} \\ \bottomrule
\end{tabular}
\end{table*}
%REFERRED TO IN Section 6.2 first para

Classifying search actions was more difficult than predicting speech acts. 
While speech acts are determined using the speaker's utterances, search actions (by the agent) are triggered by the user's utterances. 
Therefore, our goal was to connect the agent's search strategy to the user's search intent. 
The user utterance triggers the search actions but provides little contextual information about the kind of search the system should perform to solve the information problem. 
The previous utterances by the user were used to generate the linguistic features and word embeddings for BERT. 
Table \ref{tab:convex-scs-search} reports the accuracy of the models for search action prediction (for both datasets). For the CONVEX dataset, the best-performing model for classifying search action was \textit{meta+linguistic}, which had the highest accuracy of 58.8\% and median accuracy of 52.5\%. The second-best model combined all three feature categories (\textit{meta+linguistic+bert}) while the third-best used metadata features only.
The performances of all three models were comparable. 
The metadata features were most crucial for search action classification. Since we used the previously predicted speech act as a metadata feature, it improved the system's understanding of the user intention and led to a better prediction of the search strategy. The number of training instances for search action prediction was much lower than for speech act prediction. The small training data could be a reason why the BERT model and the linguistic features model could not discover the underlying patterns related to the search action, which led to the poor performance of both the individual channels and their combinations.

Once again, the performance of the models was verified on the SCS dataset, and the results are listed in Table \ref{tab:convex-scs-search}. Non-parametric Wilcoxon signed-rank test was used to assess pairwise significance.
The results validated our previous assertion about metadata features being the most important feature category for predicting search actions. The best-performing model was the single feature category \textit{meta} with the highest and median accuracy of 61.11\% and 54.44\%, respectively.
The second-best performing model was \textit{meta+linguistic}, followed by \textit{meta+bert}. For SCS, the difference between the \textit{meta} and the rest of the models was significant. The \textit{meta} model outperformed the \textit{meta+linguistic} model by 6.7 points for maximum accuracy and 8.89 points for median accuracy.

\section{Conclusion and Future Work}

This study aims to improve the natural language understanding and search performance of voice-based conversational search systems, which would, in turn, increase the adoption of such systems in digital libraries. Conversational search systems facilitate natural language queries and could improve the user experience by supporting contextual and complex queries and personalized recommendations in digital libraries.

We developed a WoZ methodology to collect user-system interaction data -- with voice-based conversations and realistic agent responses. 
Next, we revised the COR model to develop a set of themes for speech acts and search actions. To create our gold standard dataset (CONVEX), we annotated each utterance and action with the corresponding speech and search themes. 
Finally, we used an attention-based deep neural network with three different data channels to classify speech acts and then the search actions, thereby showing that with a better understanding of user utterances (intentions), a conversational system can provide better search results.
We show that using different categories of carefully engineered features makes it possible to work with smaller user study datasets and produce comparable results. Also, we have demonstrated that speech act classification on smaller user study data can be effectively performed using a multi-channel network. The multi-channel network also outperforms a fine-tuned DistilBERT model. We validate our results with a second publicly available dataset -- as a proof of concept -- to show that the results can be extended to other datasets in the future. Also, by utilizing the concept of speech acts to predict system-level search actions, we can improve the natural language understanding of future voice-based search systems.

Our study has a few limitations, which we will address in future research. While the sample size in our data is reasonable, with 25 users and 75 search sessions, with 1,834 utterances and 509 system actions, it is low for search action prediction (to train the deep neural classifiers). Also, there were minority classes present in both datasets. In the future, we would like to collect more data or combine multiple publicly available datasets. We would also like to balance the dataset through oversampling techniques like SMOTE.

Overall, this research contributes to the theory and algorithmic development of conversational search systems. Our automatic classification model can be easily extended to most user-system conversational search dialogues.
The rich conversational dataset (CONVEX) solves the need for realistic user-system information-seeking dialogues and could be used in future research. 
The CONVEX dataset and the code are available publicly. 
The findings of this paper -- validated by the publicly available SCS dataset -- provide insights on the challenges of the task, on how to better understand the user dialogues in an information-seeking conversation, and lays out the scope for other researchers to inspect and deduce new findings. %In future, we also intend to develop a working prototype of the system.

\bibliographystyle{ACM-Reference-Format}
\bibliography{reference}

\end{document}